\documentclass[12pt]{article}
\usepackage{epsfig}
\begin{document}
%set bwid 6; set fwid 6; set hwid 6; set pwid 6; set xwid 8; set ywid 6;
%igset lwid 6

\begin {center}
{\bf Comment on `Systematics of radial and angular-momentum
Regge trajectories of light non-strange $q\bar q$ states'}
\vskip 5mm
{\rm D.V.Bugg }
{\normalsize  \it Queen Mary, University of London, London
E1\,4NS, UK}
\\ [3mm]
\end {center}
\begin{abstract}
%\noindent
Masjuan, Arriola, and Broniowski [Phys. Rev. D85, 094006 (2012)]
claim that the slope of the light-quark radial  
trajectories is $1.35 \pm 0.04$ GeV$^2$, disagreeing with the Crystal 
Barrel value $1.143 \pm 0.013$ GeV$^2$.
There are defects in their choice of data.
When these defects are revised, results come back close to
the Crystal Barrel average for the slope.
A revised average value is given here.

 %{Keywords: Spectroscopy, Light Mesons}
\end {abstract}

\section {Introduction}
Masjuan, Arriola and Broniowski (referred to later as MAB for
brevity) report a new analysis of the slopes of trajectories
\cite {MAB1} \cite {MAB2}.
They adopt a $\chi^2$ criterion which increases the errors assigned
to resonance masses by up to a factor 20.
They take 
\begin {equation}
\chi^2 = \sum _n \left( \frac {M^2_n - M^2_{n,exp}}{\Gamma_nM_n}\right)^2,
\end {equation}
where $M_n$ and $\Gamma_n$ are masses and widths of fitted resonances.
Their rationale is that the extrapolation to poles off the real $s$-axis
may be inaccurate by one half-width.

The essential point of disagreement with MAB is whether this assumption is
justified or not.
My point is that large uncertainties in the extrapolation to the pole arise
only when a strong $S$-wave threshold opens in the immediate vicinity of
the resonance.
An example is $f_2(1565)$ which decays strongly to the $\omega \omega$ S-wave,
opening precisely at 1565 MeV.
In this case, there is a large dispersive contribution to the real part of
the amplitude, as discussed below in Section III.
Clearly one should be alert to such thresholds.
However, most high mass resonances have many open channels and such effects
are small.
After eliminating the special cases, there is no support for the assumption
of Eq. (1) as the general case.

To illustrate the effect of the increased errors, it is sufficient to
quote one example.
The Particle Data Group (PDG) quotes a mass for the $a_4(2040)$ as 
$2001 \pm 10$ MeV \cite {PDG}.
For $a_4(2255)$ the PDG uses two measurements with masses $2237 \pm 5$
and $2255 \pm 40$ MeV.
From these two states, MAB find a slope of $1.0 \pm 0.8$ GeV$^2$.
This implies an error in the mass difference of 203 MeV.
This is entirely inconsistent with Crystal Barrel assessment of errors
for the analytic continuation to the pole.

Further disagreements arise from several sources.
There do exist some missing states, and that must be
realised in drawing trajectories.
Another point is that it is well known that $c\bar c$ and $b\bar b$
$^3S_1$ ground states are anomalously low in mass compared 
with a straight trajectory through $J/\psi$ and $\Upsilon$ $n=2,3,4$ radial
states.
There are indications that this is also true for $n\bar n$ states.
A third point is that there is almost certainly a $J^{PC} = 0^{++}$ 
glueball in the mass range 1370 to 1800 MeV, but no present agreement
on its identification.
It will certainly mix with $q\bar q$ states, and this mixing makes
the masses of $q \bar q$ components uncertain.

A primary problem is that MAB do not distinguish clearly between 
$^3S_1$ and $^3D_1$ states.
Conventional wisdom is that $P$-state light mesons appear at masses
1200--1300 MeV, $D$ states at 1600--1700 MeV and F states near 2000 MeV. 
However, they assign the third $^3S_1$ $\omega$ state a mass of
$\sim 1970$ MeV, i.e. $\sim 300$ MeV above the $^3D_3 \, \omega_3(1670)$ 
and $\rho_3(1690)$.
The result is a slope for the $^3S_1$ $\omega$ trajectory $\sim 4/3$ times 
larger than other trajectories.

They also replace the best determinations of trajectories for $I=0$, $C=+1$ 
mesons (where there are 10 sets of data) by poorer determinations of 
$I=1$ $C=+1$ mesons, where there are no polarisation data to separate 
$^3P_2$ and $^3F_2$ mesons, hence much larger errors for masses.
They also replace the Crystal Barrel determination of the mass of
the $f_3(2300)$ by including a possibly biased mass determination
from data on $\bar pp \to \Lambda \bar \Lambda$.
That is not a good idea, since mixing with the $s\bar s$ amplitude can
confuse the situation.

In order to present the discrepancies with the slopes of trajectories
assigned by Crystal Barrel (CB), the slopes of 
all trajectories are redetermined here from final CB data sets and 
tabulated for comparison with the slopes of MAB.
The table of results makes the differences immediately apparent.

\section {Prologue}
The Crystal Barrel has produced extensive data on formation
of high mass mesons in the process $\bar pp \to R \to A + B$,
where $R$ stands for a resonance and there are 18 channels of
all-neutral final states available.
Ref. \cite {survey} reviews the data and technical details.
The detector covers $98\%$ of the solid angle with caesium 
iodide crystals which measure all-neutral final states.
Quantum numbers fall into four non-interfering families
$I=0$ or 1, $C = +1$ or -1.

For $I=0$, $C = +1$, there are data on 6 channels: $\eta \pi ^0\pi ^0$,
$\eta '\pi ^0 \pi ^0$, $3\eta$, $\pi ^0 \pi ^0$, $\eta \eta$ and
$\eta \eta'$.
There are also differential cross sections and polarisation data 
for $\bar pp \to \pi ^+\pi ^-$ from the PS172 \cite {PS172} and an 
earlier experiment at the CERN PS \cite {Eisenhandler}.
These are vital for two reasons.
First, they separate $^3P_2$ and $^3F_2$ states. 
Secondly, polarisation is phase sensitive and reduces errors 
on fitted masses and widths substantially.
The improvement in mass determination from polarisation data can
be up to a factor 4 because of its phase sensitivity.

In the $\eta \pi \pi$ data there are prominent $f_4(2050)$ and
$f_4(2300)$ signals, easily identified from their strong angular
dependence. 
They  are determined accurately in mass and width from data at 9 beam 
momenta from 600 to 1940 MeV/c.
These states serve as interferometers for all lower spin triplet 
partial waves.
There is also a lucky break, that two singlet states also appear
prominently: an $\eta_2(2250)$ in $\eta '\pi \pi$ and $\eta (2320)$
in the $3\eta$ data in the channel $f_0(1500)\eta$.
A complete set of $n\bar n$ states appears in
two towers of resonances centred near 2000 and 2270 MeV.
For $I=1$, $C=-1$, an almost complete set of states also appears,
but with poor identification of $^3S_1$ states.
For  $I=1$, $C = +1$, there are actually two solutions, with one of them 
close to the $I=0$, $C =+1$ solution, as one would expect for light quarks 
with small mass differences.
For $I=0$, $C=-1$, statistics are low for $\omega \eta$ and the
$\omega \pi ^0 \pi ^0$ data have the problem that the broad 
$\sigma \equiv f_0(500)$ interferes all over the Dalitz plot. 

The $F$ states lie systematically $\sim 70$ MeV above the $P$ states,
because high $L$ states need to overcome a centrifugal barrier in
order to resonate.
The $D$ states lie roughly midway; $S$ and $G$ states continue the
sequence.

The $f_2(1525)$ is widely accepted as the $s\bar s$ partner of
$f_2(1270)$.
Production of $f_2(1525)$ in the Crystal Barrel experiment is
extremely weak.
It is detected at the 1--2$\%$ level in $\bar pp \to \eta \eta \pi^0$
in flight \cite {Amsler}.
The conclusion is that $\bar pp$ annihilation is dominantly to
$n\bar n$ final states - hardly a surprise.
This conclusion is supported and quantified by a combined  analysis of data
on $\bar pp \to \pi^+\pi ^-$, $\pi ^0\pi ^0$, $\eta \eta$ and $\eta \eta '$
\cite {flavour}.
Amplitudes for decay to $\eta \eta$ and $\eta \eta '$ depend on the
well known composition of $\eta$ and $\eta '$ in terms of singlet
and octet states and the pseudoscalar mixing angle.
The observed state $R$ is expressed as a linear composition
$R = \cos \Phi |q \bar q> +\sin \Phi |s \bar s>$.
The result is that $\Phi \le 15^\circ$, i.e. a maximum of $25\%$ in 
amplitude, for all observed states with the exception of $f_0(2105)$ 
(which is taken as a glueball candidate, but could possibly be due to
unexpectedly strong mixing between closely spaced $n\bar n$ and 
$s\bar s$ states).
The allocation MAB make between $n\bar n$ and $s\bar s$ states is in 
conflict with the fact that CB states are dominantly $n\bar n$.

The partial wave analysis of CB data is documented in Section 4 of Ref.
\cite {survey}.
This describes systematic checks which have been made on the identification
of resonances, particularly their stability as the number of fitted
resonances was changed.
The following sections illustrate the result and discuss 
individual resonances and their Argand diagrams.
For $I=0$, $C = +1$, all states have statistical significance $>25$ standard
deviations except for the $f_2(2001)$, which is $18\sigma$ but observed
clearly in four sets of data.
Two states, $f_1(2310)$ and $\eta (2010)$ have rather large errors for masses.
Regge trajectories are discussed in section 9.
For channels $\omega \pi $ and $\omega \eta$, polarisations of $\omega$ are
determined by the angular dependence of decays to $\pi ^+\pi ^-\pi ^0$ and
are very revealing.
The interpretation of this polarisation is important and discussed in
Sections 7.1 and 8.
There is one non-standard piece of nomenclature.
These polarisations are described as vector polarisation $P_y$.
Strictly, the standard nomenclature is that this should be called
${\rm Re}\, iT_{11}$, where $T$ is tensor polarisation.

\section {Determination of slopes of trajectories}
One should be aware in advance that states may deviate from
straight trajectories because of dispersive effects on resonance
masses. 
The strict form for the denominator of a  Breit-Wigner  
amplitude is
\begin {eqnarray}
D(s) &=& M^2 - s - m(s) - i\sum _j {g^2_j \rho_j} \\
m(s) &=& \frac {1}{\pi}P\, \int ^\infty _{sthr} \sum_j
\frac {g^2_j \rho_j(s')\, ds'}{s' - s}.  
\end {eqnarray}
To make the Principal Value integral converge better, it is 
typical to make a subtraction on the resonance, although in 
principle this can be done
at any mass; $sthr$ is the $s$ value at threshold for each channel. 
The $g^2_j$ are coupling constants to every decay channel, and
$\rho_j(s')$ are the phase space for each final state, including
centrifugal barriers and possible form factors.
Near the thresholds of important decay channels, a change in the 
imaginary part of the amplitude is accompanied by a corresponding real 
part so as to obey analyticity.
At sharp thresholds, the imaginary part of the phase space rises
linearly from threshold, and produces a cusp in the real part of
the amplitude. 
This acts as an attractor \cite {sync}.
The $a_0(980)$ and $f_0(980)$ are attracted to the $K\bar K$ S-wave
threshold. 
Likewise the $f_2(1565)$ is attracted to the sharp
threshold for the $\omega \omega$ final state;
this attraction is augmented by a broader threshold in the
$\rho \rho$ channel, which has 3 times more events from the SU(2)
relation with $\omega \omega$.
The PDG mass is taken from the $\pi\pi$ channel, but the 
$\omega\omega$ and $\rho\rho$ channels are respectively a 
factor of 8.5 and a factor of 25 stronger than $\pi\pi$ when 
integrated over the available mass range, see Fig. 5(b) of 
Baker et al \cite {Baker}. 
In that paper, a dispersion relation was evaluated 
for the effect of both these channels; the pole position was 
determined to be $1598 \pm 11 (stat) \pm 9(syst)$ MeV. 
This is lower than the $a_2(1700)$ of the PDG, but this is 
because of physics which is understood.
%Fig. 1 
\begin{figure}[htb]
 \begin{center}
 \vskip -11.5mm
\includegraphics[width=12cm]{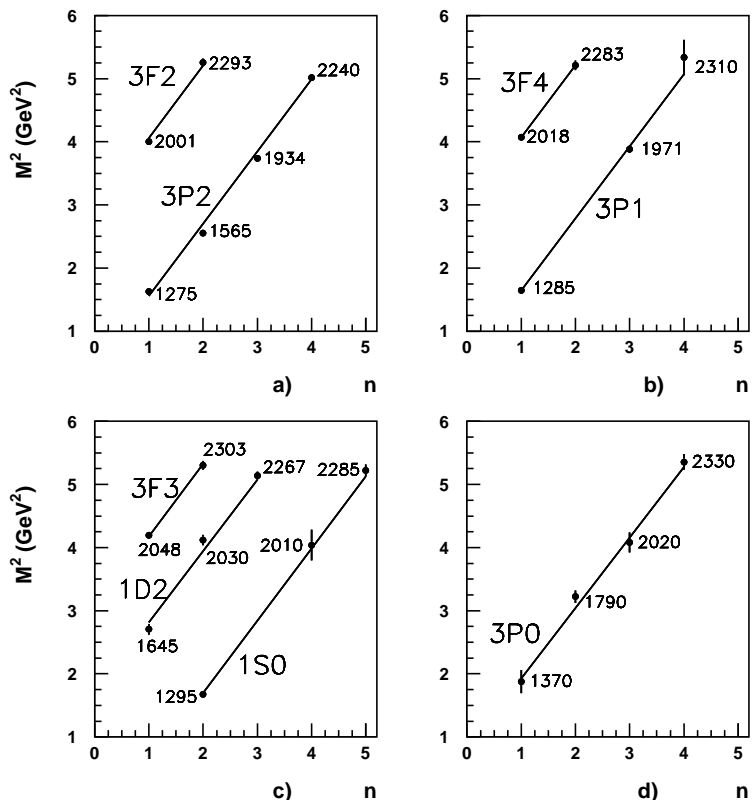}
 \vskip -8mm
 \caption {Trajectories of light mesons with $I=0$, $C=+1$ observed in
 Crystal Barrel data in flight, plotted against radial excitation number
 $n$; masses are marked in MeV. The average slope is adopted for all cases, so as
to display discrepancies. }
 \end{center}

 \end{figure}

For CB data in flight above the $\bar pp$ threshold, there is 
a conspicuous $f_2(1270)\eta'$ signal fitted as an $I=0$, 
$C=+1$ $J^{PC}=2^{-+}$ resonance at 2267 MeV and clearly 
associated with the S-wave threshold effect, see Figs. 17 and 
18 of Ref. \cite {survey}. 
For $I=1$, $C=+1$, there are signs of similar activity near 
the $a_2(1320)\eta'$ threshold, but this channel cannot be 
reconstructed accurately enough in the very difficult final 
state $\eta'\eta'\pi^0$.
S-wave structure could arise at thresholds for 
$a_2(1320)\omega$, $a_2(1320)\rho$,
$f_2(1270)\omega$ and $f_2(1270)\rho$, but is blurred out by 
convolution of the large widths of $a_2$ and $f_2$ with the 
$\rho$.
It would lead to structure distributed over $J^{PC}=3^{--}$, $2^{--}$ and $1^{--}$,
but in the absence of polarisation data cannot be sorted out at present.  
For P-wave thresholds, the imaginary part of the amplitude increases
as the cube of the momentum, and leads to negligible effects.

The $\bar pp$ and $\bar pn$ total cross sections follow a $1/v$ variation, 
where $v$ is the relativistic velocity of the incident $\bar p$.
The result is a strong cusp at the $\bar pp$ threshold in both
$^1S_0$ and $^3S_1$ partial waves.
This $1/v$ dependence is included into the partial wave analysis
for these waves.
This cusp will perturb masses of resonances near the threshold.
Data from Novosibirsk on $\bar pp \to 6\pi$ final states have a
rapid mass variation close to the $\bar pp$ threshold 
\cite {Solodov}.

Returning to the question of observed slopes of trajectories, 
each set of quantum numbers will be examined one by one, 
fitting the expected states to observed masses and errors, 
but paying particular attention to cases where states are 
missing or strongly displaced by dispersive effects.
Having done this, a grand average is taken of all slopes.
As a guide, Fig. 1 shows the updated analysis of $I=0$, 
$C=+1$ states.

\section {Results}
Table 1 lists slopes for all families of light mesons in GeV$^2$.
General comments are as follows.
First, $I=0$, $C=+1$ states are best determined, because of the
available polarisation data, which are very precise.
For triplet states, decays are possible for orbital angular
momentum $L=J+1$ and $L=J-1$.
The ratio of coupling constants $r_J = g_{J+1}/g_{J-1}$ is tabulated in 
Ref. \cite {survey}, but not tabulated by the Particle Data Group; it is 
the basic guide to whether states have $L=J+1$ or $J-1$.

Why should $^3F_2$ $n\bar n$ states decay preferentially to $\bar pp$
$^3F_2$?
A feature of CB data is that $F$ states decay strongly to
channels with high angular momentum.
The origin of this is clearly a good overlap between wave functions of
initial and final states.
Llanes-Estrada et al. point out a formal analogy with the Frank-Condon
principle of molecular physics consistent with this interpretation
\cite {lanes}.

It is immediately clear from the errors in the table that the 
$\chi^2$ weighting used by MAB increases some errors by large 
amounts.
From the agreement in many cases between CB and MAB, it is 
also clear that remaining discrepancies should be inspected 
closely.
It is easiest to compare with results from MAB in their 
Section 3, taking them in the reverse order to the 
publication, i.e. K to A.
A clear picture of disagreements then arises step by step.

The $f_1$ and $f_3$ states are considered in MAB Sec. 3K.
The CB approach is to fit $f_1(1285)$, $f_1(1910)$ and 
$f_1(2310)$ using their errors.
The first two have small errors, with the result that the fit 
misses the weak $f_1(2310)$ by just over one standard deviation.
Neither $f_1(1420)$ nor $f_1(1510)$ is fitted well by the
CB  approach or that of MAB. 
In the CB fit, a state is expected at 1660 MeV, close to its 
isospin partner $a_1(1640)$.
The $f_1(1510)$ is naturally assigned as the $s\bar s$ 
analogue of $f_1(1285)$; the mass difference is close to that 
between $f_2(1270)$ and $f_2(1525)$.
Longacre proposes that $f_1(1420)$ is a molecule where an $
L=1$ pion circles a $K\bar K$ core \cite {Longacre}.
For $f_3$, the MAB slope has an error a factor 9 larger than 
the CB value.
Sections J and I agree well on slopes between CB and MAB for
$b_3$, $b_1$, $h_3$ and $h_1$ states.

For $f_3$, it is clear from the slope of MAB that they use a 
mass well above the CB determination of the mass of 
$f_3(2300)$, $2303 \pm 15$ MeV; it seems likely that it is 
replaced by the value $2334 \pm 25$ MeV from 
$\bar pp \to \Lambda \bar \Lambda$ data.
This is dangerous, since it may introduce mixing with $s\bar s$.
If one takes the weighted mean of the two masses quoted by the 
PDG \cite {PDG} for $f_3(2300)$, namely $2311 \pm 13$ MeV, the 
slope is $1.15 \pm 0.08$.

\newpage
\begin{table}[h]
\begin{center}
\begin{tabular}{ccccc}
\hline
Family & Nominal masses(MeV) & CB slope & MAB slope & Comments\\\hline
$f_4$    & 2050,2300 & $1.140 \pm 0.093$ & - & \\
$f_3$    & 2050,2300 & $1.147 \pm 0.071$ & $1.27 \pm 0.64$ & + \\
$f_2(^3F_2)$ & 2000,2295 & $1.254 \pm 0.075$ &  - &  \\
$f_2(^3P_2)$ & 1270,1565,1910,2240 & $1.113 \pm 0.025$  &  - & + \\
$f_1$    & 1285, -- ,1970,2310 & $1.130 \pm 0.064$ &$1.19 \pm 0.15$ & + \\
$f_0$    & 1370,--,--,2330& $1.24 \pm 0.045$ &$1.24 \pm 0.18$ & +\\
$h_3$    & 2025,2275 & $1.075 \pm 0.147$ & $1.08 \pm 0.54$ & \\
$h_1$    &1170,1595,1965,2215& $1.195 \pm 0.059$ & $1.20 \pm 0.25$ &  \\
$b_3$   &  2025,2275 & $0.95 \pm 0.191$   &$1.08 \pm 0.54$ & \\
$b_1$   &1235,--,1960,2240 &$1.169 \pm 0.058$ &$1.17 \pm 0.18$ \\
$\omega_3$&1670,1945,2255& $1.059 \pm 0.163$ & $1.16 \pm 0.26$ & \\
$\omega_2$&1975,2195   & $0.917 \pm 0.160$ & -  & \\
$\omega_1(^3D_1)$&1650,1960,2295& $1.091 \pm 0.068$ & $1.27 \pm 0.47$ &+\\
$\omega_1(^3S_1)$&782,1420,1650,--,2.205&$1.080\pm 0.029$ & $1.50 \pm 0.12$ &+ \\
$a_4$    &2040,2255 &$1.000 \pm 0.045$ &$1.00 \pm 0.8$ \\
$a_3$    &2030,2275 &$1.051 \pm 0.170$ &$1.5 \pm 0.11$ \\
$a_2(^3F_2)$&2030,2255  & $0.964 \pm 0.128$  & $1.00 \pm 0.7$ & \\
$a_2(^3P_2)$&1320,1700,1950,2175 & $1.000 \pm 0.060$  & $1.39 \pm 0.26$ & + \\
$a_1$    & 1260,1640,1930,2270 & $1.084 \pm 0.063$ & $1.36 \pm 0.49$ & + \\ 
$a_0$    & 1450,2025& $0.964 \pm 0.073$ & $1.42 \pm 0.26$ & + \\ 
$\pi _2$ & 1670,2005,2245 & $1.218 \pm 0.062$ & $1.21 \pm 0.36 $ & \\
$\pi _0$ & 1300,1800?,2070,2360 & $1.29 \pm 0.200$ & $1.27 \pm 0.27 $ & \\
$\rho_3$ & 1690,1990,2265 & $1.094 \pm$ 0.050& $1.19 \pm 0.32$ & + \\
$\rho_2$ & 1940,2225 & $1.206 \pm 0.085$ & - &  \\
$\rho _1(^3D_1)$ &1700,2000,2270 & $1.203 \pm 0.052$ &$1.08 \pm 0.47$ & + \\
$\rho_1(^3S_1)$ & 770,1450,--,1900,2150 & $1.365 \pm 0.108$ &$1.43 \pm 0.13$ & + \\
$\eta_2$ & 1645,2030,2265 & $1.188 \pm 0.038$ & $1.32 \pm 0.32$ & \\
$\eta_0$ & 1295,2320 & $1.241 \pm 0.030$ & $1.33 \pm 0.11$ & + \\
\hline                                             

\end {tabular}
\caption {Nominal resonances masses and slopes (in GeV$^2$) from CB and MAB 
analyses; comments are marked by a $+$ in the final column and discussed
in the text; a dash in column 2 indicates a missing state
(or unused for reasons discussed in the text).}
\end{center}
\end{table}

Sections 3J and 3I of Ref. \cite {MAB1} agree between CB and 
MAB for $b_3$, $b_1$, $h_3$ and $h_1$ states.
Section 3H considers $\omega$ states.
MAB and CB agree on the $\omega_3$ trajectory within
errors.
But MAB put the $\omega(1650)$ on a trajectory with an 
$\omega (2290)$ observed in $\bar pp \to \Lambda 
\bar {\Lambda}$.
This could be an $s\bar s$ state or could be due to an
upward shift of the $\omega (2255)$ because of $s\bar s$ 
background.
The consequence is that MAB ignore the well known 
$\omega (1650)$ as a standard $^3D_1$ state.

The scheme listed in the PDG meson summary tables takes the 
$\omega (1420)$ as a $^3S_1$ state and $\omega (1650)$ as 
$^3D_1$.
The $\rho$ states run parallel, with $\rho(1450)$ as a $^3S_1$ 
state and the well established $\rho_1(1700)$ as $^3D_1$.
The $\phi (1680)$ fits in as $^3S_1$ and $\phi (1850)$ as 
$^3D_1$; the $\phi$ states lie higher than $n\bar n$ states 
by a similar mass difference to that between $f_2(1270)$ and 
$f_2(1525)$ (which is well established as $s\bar s$).
Fitting the $\omega (1960)$ together with $\omega (1650)$ 
gives a slope of $1.091 \pm 0.068$, close to other CB slopes.
Fitting the remaining $\omega (^3S_1)$ trajectory with states 
at 782,1425 and 2205 MeV, but with two missing states gives a 
slope of $1.080 \pm 0.029$. 
The fit to the second state however gives it a mass of 1329 MeV.
Many authors have noticed this point. 
It could well be due to the fact that the ground state is 
abnormally low in mass, just as the $J/\psi$ and 
$\Upsilon (1S)$ lie significantly below a 
straight line through $n=2$, 3 and 4 radial excitations. 

Section  3G of Ref. \cite {MAB1} considers $f_0$ states.
Here physics remarks are required.
One of the fundamentals of Particle Physics is chiral symmetry
breaking.
This was first proposed by Gell-Mann and L\' evy in 1960 \cite {Levy}.
It became well known to theorists that the attracive $NN$ interaction
requires exchange of two correlated pions with a broad peak at
450--650 MeV, denoted by the $\sigma$.
Bicudo and Ribiero provided a detailed account of how this arises \cite {Bicudo}.
The mechanism today accounts for the low mass 
$\sigma \equiv f_0(500)$, $\kappa$, $a_0(980)$ and $f_0(980)$.
It is now well understood \cite {Weise} how a crossover arises
between these exceptional states and regular $q\bar q$ states 
near 1 GeV.
The surviving mixing above 1 GeV is likely to push the $q\bar q $
$0^+$ states up in mass. 
This can explain the anomalously high masses of $f_0(1370)$ and
$a_0(1450)$.
A further complication for $f_0$ is the likely existence of a
glueball in the mass range 1500--1800 MeV, still obscure.
A further point is that there is evidence \cite {1790} 
that the $f_2(1810)$ claimed  by GAMS has been confused with the
$f_0(1790)$ candidate for the radial excitation of $f_0(1370)$; 
the $f_0(1790)$ is consistent with the BES II $0^+$ peak 
observed at 1812 MeV in $\omega \phi$ decays\cite {1812}.
So, in summary, $f_0$ states are complex.
Conclusions about the $f_0$ slope are therefore ambiguous.
Using only the $\pi \pi$ mass of $f_0(1370)$ and the mass of 
$f_0(2330)$, the slope is $1.24 \pm 0.045$ GeV$^2$.

The $f_0(1370)$ decays dominantly to $4\pi$; this introduces 
large dispersive effects on the mass.
Crystal Barrel data at rest on $\bar pp \to 3\pi ^0$ contain 
600,000 precisely measured events and interference effects 
between the three $\pi\pi$ components determine phase 
variations very precisely. 
The mass fitted to the $\pi \pi$ channel is 
$1309 \pm 1(stat) \pm 15(syst)$ MeV.
However rapidly increasing phase space for its dominant 
$\rho \rho$ decay channel moves the peak in $4\pi$ data up 
by $\sim 75$ MeV. 
Pole positions on different sheets are given in Table 4 of 
\cite {1370} for $f_0(1370)$ and $f_0(1500)$ and are quite 
revealing.
This strong threshold shifts the pole of $f_0(1370)$ by at 
most 17 MeV from the peak in $\pi \pi$; the fitted 
$2\pi$ full-width is 325 MeV.
So in this extreme case, the shift in pole position is 
$\le 10.5\%$ of the half-width.
For the $f_0(1500)$, the shift in pole position is 8 MeV from 
the nominal mass compared with a half-width of $54.4 \pm 3.5$ 
MeV.
These shifts are a factor at least 6 less than MAB assume.
 
MAB construct two trajectories for what they take to be 
$n\bar n$ states and $s\bar s$.
The $n\bar n$ trajectory starts with $f_0(980)$ and finishes 
with $f_0(2200)$.
However, there is almost universal agreement today that 
$f_0(980)$ and $a_0(980)$ are not $q\bar q$ states but have 
dominant 4-quark composition \cite {Achasov}. 
The $f_0(980)$ decays dominantly to $K\bar K$, not $\pi \pi$.
BES II quote a  $KK/\pi\pi$ branching ratio  
$4.21 \pm 0.2(stat) \pm 0.21(syst)$ \cite {Dong}; this is one of the 
very few experiments which has data on both $K\bar K$ and $\pi \pi$.
A further important source of information is the decay of $\chi^0 \to
K^+K^-$.
There is a conspicuous $f_0(1710)$ in the data and a further peak at
$f_0(2200)$ \cite {chi0}.
This suggests that they both have substantial $s\bar s$ components.
So the MAB $n\bar n$ trajectory looks unlikely.

On their $s\bar s$ trajectory, MAB start with $f_0(1370)$ and finish
with $f_0(2330)$.
The $f_0(1370)$ is observed dominantly in decays to $4\pi$, largely 
$\rho \rho$.
It has a branching ratio to $K\bar K$ of $0.12 \pm 0.06$ in CB data, 
so it does not look like an $s \bar s$ state. 
The last member of this trajectory is $f_0(2330)$.
This has been observed in Crystal Barrel data in decays to
$\pi \pi$ and $\eta \eta$  \cite {flavour} with a flavour
angle of $15.1^\circ$, so it is certainly not a dominantly
$s\bar s$ state.

Section E of MAB discusses $a_0$, $a_2$ and $a_4$ states.
This is again a complex story. 
They use the mass of $a_0(980)$, which is unwise in view of its
association with chiral symmetry breaking. 
The $a_0(1450)$ is well determined \cite {1450}, but the 
signal for $a_0(2025)$ is very weak.
An additional $a_0$ is to be expected somewhere between
these two, but there are no data adequate to detect
it; finding spin 0 states is difficult. 
Assuming this state has been missed, the slope from $a_0(1450)$
and $a_0(2025)$ is $0.96 \pm 0.08$ GeV$^2$, but is probably
affected by chiral symmetry breaking and is not used in the
overall CB average for the slope. 
MAB make the opposite assumption that this is the first radial
excitation of $a_0(1450)$ and hence find a slope of 
$1.42 \pm 0.26$.

Moving on to $a_2$ states, MAB consider as an upper trajectory
($^3F_2$) $a_2(2030)$ and $a_2(2255)$ and
arrive at a similar slope to CB.
For the $^3P_2$ trajectory, they take $a_2(1320)$, $a_2(1700)$ 
and $a_2(2175)$.
This is to be compared with the well established $f_2$ 
trajectory $f_2(1270)$, $f_2(1565)$, $f_2(1910)$, $f_2(2240)$.
They miss an $a_2$ state to be identified
with the $a_2(1950)$ of Anisovich et al. \cite {01F}.
They find a slope $1.39 \pm 0.26$ compared with the CB 
determination $1.00 \pm 0.06$.

There is agreement between CB and MAB for the $a_4$ slope.
However, one comment is needed on the PDG determination of the
mass.
It is determined largely by the data of Uman et al. 
\cite {Uman}.
If one looks at their Fig. 6, the difference in $\chi^2$ between
$a_2$ and $a_4$ is small.
They do not consider the possibility that both $a_2$ and $a_4$
are present; that would not be at all surprising.
Therefore the CB determination of the mass is preferred here.

Consider next $f_2$ states.
The problem here is that MAB do not discriminate between the
$^3P_2$ states and $^3F_2$, which are well separated in CB data.
MAB launch into four alternative scenarios, all of which
have problems.

Their $f_2^a$ trajectory is made from $f_2(1370)$, $f_2(1750)$
and $f_2(2150)$.
The $f_2(2150)$ is not seen in CB data.
It is observed only in decays to $K\bar K$ and $\eta \eta$.
The $f_2(2010)$ of the PDG \cite {PDG}, observed by 
Etkin et al. in $\bar pp \to \phi \phi$ actually peaks at 
2150 MeV.
The mass quoted by Etkin et al. \cite {Etkin} is the
K-matrix mass, and can differ from the T-matrix mass;
the K-matrix formalism assumes that all decay channels are known,
but that is unlikely.
The obvious interpretation of the $f_2(2150)$ is the $s\bar s$
partner of $f_2(1910)$, i.e. a $^3P_2$ state.
Etkin et al. also report an $f_2(2300)$ in the $\phi \phi$ S-wave
and $f_2(2340)$ in the $\phi \phi$ D-wave.
This is naturally to be interpreted as an $s\bar s$ $^3F_2$ state.
The $f_2(1750)$ of Schegelsky et al. \cite {Schegelsky} is observed 
in $\gamma \gamma \to K\bar K$, and is interpreted by them as an 
$s\bar s$ state - the radial excitation of $f_2(1525)$, though 
rather low in mass.

The MAB $f_2^b$ trajectory uses $f_2(1430)$.
That entry in PDG tables has a straightforward interpretation.
The $\omega \omega $ channel (and therefore $\rho \rho$) couples
strongly to $f_2(1565)$.
When analysing data on Dalitz plots, it is necessary to continue
the Flatt\' e formula for $f_2(1565)$ below the $\omega \omega$ 
threshold, rather than just cutting it off. 
This is the way in which Crystal Barrel analyses Dalitz plots.
The result is a cusp in the $\pi \pi$ channel at
1430 MeV, see Fig. 7 of Adomeit et al. \cite {1430}.
It is likely that the data listed by the PDG under $f_0(1430)$
were due to this phenomenon.

The MAB $f_2^c$ trajectory uses $f_2(1525)$, which is a
well known $s\bar s$ state and is obviously invalid.
Their $f_2^d$ trajectory uses $f_2(1565)$, $f_2(2000)$ and
$f_2(2295)$, hence mixing $^3P_2$ and $^3F_2$ states.
This is also invalid.

They continue with three further trajectories, the first based
on $f_2(1640)$ together with $f_2(2150)$.
The $f_2(1640)$ has been explained by Baker et al. \cite {Baker}
as the $\omega \omega$ decay of $f_2(1565)$; 
the rapidly rising $\omega \omega$ phase space shifts the peak 
in $\omega \omega$ up to 1640 MeV \cite {Baker}.
The second is based on the questionable $f_2(1810)$ and
$f_2(2220)$, which is a very narrow peak, 23 MeV wide, claimed
in BES II data.
If such a narrow state contributes to non-strange $q\bar q$ states,
it is a mystery why it is not observed very conspicuously in
CB data.
Their final trajectory is made of $f_2 (2010)$ and $f_2(2340)$,
which are observed in $\phi \phi$ and $K\bar K$
and finds a slope of $1.43 \pm 0.83$ GeV$^2$; they are
obvious candidates for $s\bar s$ states.

Section 3D of MAB \cite {MAB1} concerns $\pi$ and 
$\pi _2$ trajectories.
These agree with CB.
The $\pi (1800)$ is usually considered as a hybrid candidate; 
it has little effect on the fitted slope.

Section C discusses $\rho_1$ and $\rho_3$ states.
Their slope for the latter is close to the CB value
but with much larger error from their $\chi^2$ criterion.
The physics situation concerning $\rho_1$ states is a mess, for 
physics reasons.
The $\rho(1450)$ couples weakly to $2\pi$ and there are large
dispersive effects in the $4\pi$ channel, which have not yet
been taken into account. 
The natural interpretation of it is the $^3S_1$ radial excitation
of $\rho (770)$, but the large slope may arise simply from the
fact that the ground state is abnormally low, like the 
$J/\psi$ and $\Upsilon (1S)$.
The $\rho (1570)$ of Babar has a larger error in mass: 
$\pm 36(stat) \pm 62(syst)$ MeV and is marginally consistent
with $\rho (1450)$, which actually has a mass of $1465 \pm 25$ MeV.

The $\rho (1900)$ can be identified with a recent Novosibirsk
observation of a $6\pi$ peak almost exactly at the $\bar pp$
threshold \cite {Solodov}. 
This is likely to be a $^3S_1$ state captured by the very strong
$\bar pp$ S-wave, but could  be a non-resonant cusp.

CB data list ratios $r_J$ of coupling constants to orbital angular
momentum $L = J+1$ and $L=J-1$.
The $\rho (2000)$ has a sizable $r$ value $0.70 \pm 0.23$ 
requiring at least some $^3D_1$ contribution. 
The $\rho (2150)$ in CB data has an $r$ value $-0.05 \pm 0.42$,
consistent with $^3S_1$.
The $\rho (2270)$ in CB data has $r = -0.55 \pm 0.66$ which is ambiguous.
However, the $\rho(2000),(2150),(2270)$ make a natural sequence of
$^3D_1$, $^3S_1$, $^3D_1$.
If this solution to the puzzle is accepted, the slope of
the CB $^3S_1$ trajectory is $1.34 \pm 0.03$, with a rather poor fit to
the mass of $\rho (1450)$.
In view of the problems with $\rho (1450)$, this is not included in
the grand average.
The trajectory of $\rho (1700)$, $\rho (2000)$ and $\rho (2270)$
gives a slope of $1.17 \pm 0.03$; MAB quote $1.08 \pm 0.47$. 

Section 3B of MAB \cite {MAB1} discusses $\eta$ and 
$\eta_2$ states.
The $\eta (548)$ is believed to be a Goldstone boson and should
not be included in the assessement of $q\bar q$ states.
The $\eta (1760)$ and $\eta (2100)$ were claimed by DM2, but
later identified in Mark III data \cite {MarkIII} as having 
$J^{PC}=0^{++}$, though they sit on a large non-interfering $0^{-+}$ 
background; [$0^{++}$ and $0^{-+}$ do not interfere in $J/\psi$ 
radiative decays after summing over relevant spin states of 
the $J/\psi$]. 
They are also identified in E760 data \cite {E760} in the 
$\eta \eta$ channel, where $J^{PC} = 0^{-+}$ is forbidden by 
Bose statistics.
The remaining trajectory,  $\eta (1295)$ and $\eta (2320)$ 
gives a CB slope of $1.24 \pm 0.03$; MAB quote $1.33 \pm 0.11$.

For $\eta_2$ states, the averaged slope agrees with the global
average within errors but the $\chi^2$ of the fit to the 
$\eta_2(2030)$ in the middle is high.
There is a likely explanation.
There is an extra state $\eta _2(1870)$ which is naturally explained as
a hybrid partner to $\pi_1(1600)$ predicted near this mass. 
By the usual level repulsion, this pushes the $\eta _2(1645)$ 
down and the $\eta_2(2030)$ up, though the overall effect on 
the average slope is small.

Section 3A of MAB \cite {MAB1} discusses the $a_1$ trajectory.
The $a_1(2095)$ has a large error in mass of $\pm 121$ MeV.
The $a_1(2270)$ completes the trajectory.
There is an obvious problem that the $a_1(1260)$ has a large 
width; the PDG quotes it as 250--660 MeV. 
A recent Babar estimate is $410 \pm 31 \pm 30$ MeV.
The CB slope is $1.08 \pm 0.06$ GeV$^2$; MAB find 
$1.43 \pm 0.26$ GeV$^2$.

Finally, MAB discard all slope determinations which have only 
two points.
This removes all the determinations from $^3F_2$, $^3F_3$ and
$^3F_4$ states which are amongst the best.
As one sees from Table 1 and Fig. 1, these determinations have 
slopes consistent with other CB values.
MAB also assign $a_2(2030)$ and $a_2(2255)$ the radial quantum 
numbers $n=2$ and 3, while the nearby $f_2(2000)$ and 
$f_4(2295)$ obviously have $n=3$ and 4. 

In summary, the MAB classification of slopes unfortunately 
contains a number of problems, and there is no significant 
case for the large slopes they claim for some cases.

\section {Epilogue}
Values of CB and MAB differ significantly only where there are 
clear problems in their selection of states in the fit.
The weighted mean of CB slopes is revised slightly.
On close inspection, the $\chi^2$ contributions from 
$^3S_1\, \rho_1$ results are high by a factor 4.
Warnings about the problem in this case have already been given.
Likewise contributions to $\chi^2$ from $a_0$ are high by  a
factor 5. 
Again the text has pointed out problems for these states.
Finally, $\chi^2 $ contributions from $a_2$ $^3P_2$ states are 
high by a factor 4.
This is no surprise, since there are no polarisation data to 
provide clear identifications of these states.

MAB remark that Anisovich, Anisovich and Sarantsev (AAS) proposed a
scheme in the year 2000 where the lowest $J^{PC} = 0^{++}$ 
states were taken as $a_0(980)$ and $f_0(980)$ \cite {AAS}.
Since then, there have been many studies of the effects of chiral
symmetry breaking. 
It is now widely believed that the $\sigma$, $\kappa$, $a_0(980)$ and
$f_0(980)$ are meson-meson states, and that there is a crossover to
$q\bar q$ states near 1 GeV, where chiral symmetry breaking 
decreases rapidly.

Summarising, the mean slope without any corrections is $1.130 \pm 0.011$.
Reducing the weights of the three troublesome cases to 1 modifies this to
the final value $1.135 \pm 0.012$, compared with the old value of
$1.143 \pm 0.013$ GeV$^2$. 
There is no clear case for the large error assessment of MAB.
In fact, states with large widths already enlarge the errors for masses
appropriately.

A comment on the MAB approach is that they adopt their 
assumption that meson masses can move by $\Gamma /2$ from 
theoretical predictions;
those are that for large-$N_c$ the strong coupling constant 
scales as $g \sim 1/\sqrt {N_c}$ with the result that meson 
masses change by $\Gamma /2$ when evolved from $N_c = 3$ to 
$N_c = \infty$, see the references 16-18 given in their 
paper. 
The conclusion from the present analysis is that the $N_c$ 
world is different from $N_c \to \infty$.

The PDG lists CB data under `Other Light Mesons, further 
states' on the grounds that they need confirmation. 
Perhaps, but $I=0$, $C=+1$ does contains a complete spectrum of expected 
states.
For other isospin and $C$ values, it is desirable to improve 
the data base. 
That cannot be done in production experiments, because the 
exchanged meson is not usually known, i.e. no polarisation 
information is available.
The $\rho$ and $\omega$ states can be improved at VEPP 2 in 
Novosibirsk by using transversely polarised electrons. 
Two measurements are readily made of asymmetries normal to the
plane of polarisation and in the plane of polarisation.
Electron polarisation of $70\%$ is already achieved and two detectors
CMD and SMD are available and running.  
The presence of $^3D_1$ states is then revealed by distinctive
azimuthal angular dependence in the polarisation and can measure
whether these are pure $^3D_1$ states or linear combinations 
with $^3S_1$, and if so how big the contributions are. 
Longitudinal polarisation does not help much because it depends 
only on the difference of intensities of the two helicities 
available. 

In order to trace the missing states above 1910 MeV, polarisation 
measurements are needed for $I=1$ $C=+1$ 
($\eta \eta \pi^0$, $\eta \pi^0$ and $3\pi^0$), $I=1$, $C=-1$ 
($\omega \pi^0$ and  $\eta\omega \pi^0$), and $I=0$, $C = -1$ ($\omega \eta$ 
and $\omega \pi^0\pi^0$).
Polarisation data also introduce interference between singlet and 
triplet states, hence determining the singlet states much better.
Data are required down to $\bar p$ momenta of $\sim 360$ MeV/c, the
lowest momentum reached in the PS172 experiment \cite {PS172}.
The PANDA experiment cannot do this measurement because their lowest
available beam momentum will be 1.5 GeV/c.

\end {document}